\documentclass[preprint,showpacs,preprintnumbers,amsmath,amssymb]{revtex4}
\usepackage{graphicx}
\usepackage{dcolumn}
\usepackage{bm}
\usepackage{epstopdf}%

\begin{document}


\title{Preparation and photoemission investigation of bulk-like $\alpha$-Mn films on W(110)}

\author{Yu. S. Dedkov,$^{1,}$\footnote{Corresponding author. E-mail: dedkov@fhi-berlin.mpg.de} E. N. Voloshina,$^2$ and M. Richter$^3$}
\affiliation{\mbox{$^1$Fritz-Haber Institut der Max-Planck Gesellschaft, 14195 Berlin, Germany}\\
$^2$Institut f\"ur Chemie und Biochemie - Physikalische und Theoretische Chemie, Freie Universit\"at Berlin, Takustrasse 3, 14195 Berlin, Germany\\
 \mbox{$^3$IFW Dresden e.V., P. O. Box 270\,116, 01171 Dresden, Germany}}

\date{\today}

\begin{abstract}

We report the successful stabilization of a thick bulk-like, distorted $\alpha$-Mn film with (110) orientation on a W(110) substrate. The observed $(3\times3)$ overstructure for the Mn film with respect to the original W(110) low-energy electron diffraction pattern is consistent with the presented structure model. The possibility to stabilize such a pseudomorphic Mn film is supported by density functional total energy calculations. Angle-resolved photoemission spectra of the stabilized $\alpha$-Mn(110) film show weak dispersions of the valence band electronic states in accordance with the large unit cell.

\end{abstract}


\maketitle

Manganese (Mn) can be considered as the most complex of all metallic elements from the crystallographic point of view. Assuming regular structural trends as in the series of the $4d$ and $5d$ transition metals, one would expect crystallization of Mn in a hexagonal close-packed (\textit{hcp}) structure. In the row of the $3d$ elements, the occurence of magnetism disturbs this regular structural sequence: Fe and Co crystallize in body-centered (\textit{bcc}) and \textit{hcp} structures, respectively, whereas \textit{hcp} and face-centered cubic (\textit{fcc}) structures would be expected from the group sequences~\cite{Soderlind:1994}. Being a member of the $3d$-row, Mn behaves in a completely different way. Depending on temperature and pressure, it exists in five allotropic forms~\cite{Pearson:1958,Hobbs:2003,Hafner:2003}. $\alpha$-Mn, the stable phase below 1000\,K, has an exotic \textit{bcc} crystal structure containing 58 atoms in the conventional cubic unit cell [Fig.\,1(a)]. $\beta$-Mn is stable between 1000\,K and 1368\,K and has a simple cubic structure with 20 atoms per unit cell. An \textit{fcc} $\gamma$-Mn phase exists between 1368\,K and 1406\,K and a \textit{bcc} $\delta$-phase from 1406\,K up to the melting point. An \textit{hcp} $\epsilon$-phase of Mn exists above a pressure of 165\,GPa~\cite{Fujihisa:1995, Zheng:1998}.

Mn was grown in the form of ultrathin films on a number of \textit{fcc} (Cu~\cite{Wuttig:1993a}, Ni~\cite{Wuttig:1993b}, Ir~\cite{Andrieu:1996}, Au~\cite{Fonin:2003,Dedkov:2006mnau}) and \textit{bcc} (Fe~\cite{Walker:1993,Kim:1996}, V~\cite{Tian:1999}, W~\cite{Bode:1999,Bode:2002}) metal surfaces. In case of low-temperature growth of Mn on noble and transition metal surfaces a tetragonally expanded \textit{fct} Mn layer (with antiferromagnetic ordering) is formed which can be considered as distorted $\gamma$-Mn. Deposition of Mn at higher temperatures leads to interdiffusion and to the formation of surface alloys with $c(2\times2)$ structure, where the Mn atoms are coupled ferromagnetically in the alloy layer. 

A misfit of only 2.5\% between the lattice constants of W (3.16\,\AA) and of the high-temperature \textit{bcc} phase $\delta$-Mn (3.08\,\AA\ at 1133$^\circ$\,C) opens a chance prepare relatively thick distorted $\delta$-Mn(110) films on W(110) [see Fig.\,1(b), lower part]. Recently, the possibility to stabilize thin distorted $\delta$-Mn films up to several monolayers (ML) on W(110) was confirmed by low-energy electron diffraction (LEED) and scanning tunneling microscopy (STM)~\cite{Bode:1999,Bode:2002}. The epitaxial relationship is mantained up to a coverage of about 3\,ML, but in contrast to Fe/W(110), already the first Mn layer exhibits a modest growth anisotropy in the $\langle 001 \rangle$ direction~\cite{Bode:1999}.

Concerning the study of electronic properties, all $3d$ metals except Mn were routinely investigated by means of photoelectron spectroscopy with angle- as well as spin-resolution~\cite{Hufner,Rader:1999}. This lack of photoemission (PE) data for Mn seems surprising, but might be due the complicated crystal structure of $\alpha$-Mn, although epitaxial growth of Mn overlayers would be a possible way to obtain crystalline bulk-like samples.

Here we report crystallographic and electronic structure investigations of stabilized bulk-like distorted $\alpha$-Mn films on W(110). The formation of $\alpha$-Mn was achieved by annealing of freshly \textit{in situ} evaporated Mn films at $T\approx200^\circ$\,C. LEED images show a clear $(3\times3)$ overstructure with respect to the W(110) surface that we assigned to the formation of a highly strained $\alpha$-Mn thick film with (110) orientation [see Fig.\,1(b), upper part]. Density functional (DFT) calculations confirm that this strained film has a lower total energy than the other competing Mn phases. Angle-resolved valence-band PE shows weak dispersion in agreement with the calculated band structure.

The experiments were performed in a photoelectron spectroscopy setup consisting of two chambers described in detail elsewhere~\cite{Dedkov:2006,Dedkov:2007}. A W(110) single crystal was used as a substrate. Prior to preparation of the Mn film, a well established cleaning procedure of the W-substrate was applied: repeated cycles of heating up to 1300$^\circ$\,C in an oxygen atmosphere at a pressure of $5\times10^{-8}$\,mbar and subsequent flashing up to 2300$^\circ$\,C. LEED patterns of the W substrate reveal the sharp pseudo-hexagonal structure expected for the \textit{bcc} (110) surface [see Fig. 2(b)]. Mn metal was evaporated from a carefully degassed electron-beam heated tungsten crucible. The base pressure in the experimental setup was below $5\times10^{-11}$\,mbar and during metal evaporation it was below $2\times10^{-10}$\,mbar.  The crystallographic ordering and cleanliness of the metal films were monitored by LEED and by x-ray photoelectron spectroscopy (XPS) of core levels as well as PE spectroscopy of the valence band, respectively. PE spectra were recorded at 21.2, 40.8 eV (He\,I$\alpha$, He\,II$\alpha$, UPS) and 1253.6, 1486.6 eV (Mg\,K$\alpha$, Al\,K$\alpha$, XPS) photon energies using a hemispherical energy analyzer SPECS PHOIBOS 150. The energy resolution of the analyzer was set to 50 and 500\,meV for UPS and XPS, respectively.

Fig.\,2(a) shows XPS spectra obtained during the preparation of an ordered Mn film on W(110). The emission lines corresponding to the excitations from Mn $2p$, W $4d$, and W $4f$ core levels are clearly resolved. These energy regions are marked by shadows. The series of PE peaks below 750\,eV binding energy (BE) corresponds to the Auger-electron emission lines of Mn. Core-level peaks in this figure can be used for monitoring the growth of a continuous Mn film: annealing of the freshly deposited Mn film with a thickness of 50 or 90\,\AA\ (XPS spectrum \#1) at a temperature $T\approx200^\circ$\,C for 3\,min does not change the shape of the XPS spectra [XPS spectra \#2 and \#4 in Fig.\,2(a)]. Increasing the annealing temperature to $T\approx300^\circ$\,C leads to a break of the continuous Mn film and to the formation of high Mn islands on top of the W(110)-surface covered by an ultrathin Mn film. For such an annealed sample the XPS spectra show significant W $4d$ and W $4f$ emissions in the spectra with simultaneous weakening of the Mn $2p$ emission signal. This situation is valid for any thicknesses of the predeposited Mn [see XPS spectra \#3 and \#5 in Fig.\,2(a)]. This behavior can be explained by the very high strain in the thick Mn film on W(110) due to the large misfit between the (110) planes of \textit{bcc} Mn and W (see below).

LEED images of clean W (110) and of freshly deposited 50 (90)\,\AA-thick Mn films before and after annealing at $T\approx200^\circ$\,C and $T\approx300^\circ$\,C are shown in Fig.\,2 (c), (d), and (e), respectively. Before annealing the Mn film is characterized by a uniform background with very weak spots in the LEED image indicating a high degree of disorder in the freshly deposited film [the corresponding XPS spectrum is marked by \#1 in Fig.\,2(a)]. After annealing of this  film at $T\approx200^\circ$\,C a clear $(3\times3)$ overstructure with respect to the W(110) surface is observed in the LEED image. This annealing step corresponds to the XPS spectra \#2 and \#4 in Fig.\,2(a) indicating a continuous Mn film on W(110). Annealing of the film at $T\approx300^\circ$\,C leads to the formation of a sharp $(1\times1)$ LEED structure. In this case, as mentioned before, the Mn film is broken and forms high islands characterized by the XPS spectra \#3 and \#5 in Fig.\,2(a).

The $(3\times3)$ overstructure observed for the continuous thick Mn film on W(110) can be understood on the basis of the following consideration. Since \textit{bcc} $\alpha$-Mn is the most stable phase of this polymorph metal, growth of this phase is the most probable option for a smooth thick film. In this case, Mn(110)/W(110) growth is suggested with Mn$\langle 1\overline{1}0\rangle ||$W$\langle 1\overline{1}0\rangle$ and Mn$\langle 001\rangle ||$W$\langle 001\rangle$ as shown in the upper part of Fig.\,1(b). In coincidence with the LEED data, this assumption yields lattice parameters $d_{\langle 001\rangle}^{Mn}=9.21$\,\AA\ and $d_{\langle 1\overline{1}0\rangle}^{Mn}=13.15$\,\AA\ with a related straining of the Mn lattice by $\approx3.3$\% in $\langle 001\rangle $ and by $\approx4.3$\% in $\langle 1\overline{1}0\rangle$ directions  with respect to the lattice constant of bulk $\alpha$-Mn $a_{bulk}^{Mn}=8.91$\,\AA~\cite{Pearson:1958,Yamada:1970}. Here we conclude, that a strained bulk-like $\alpha$-Mn(110) film can be stabilized on the W(110) surface via careful annealing at a relatively low temperature. Increasing the annealing temperature leads to a collapse of the metastable film and probably to the formation of a pseudomorphic thin layer of Mn on W(110) accompanied by large Mn islands. 

The electronic structure of the every-time freshly prepared Mn overlayers was studied by means of angle-resolved PE after applying the two described different annealing schemes. Fig.\,3(a) shows angle-resolved PE spectra measured with He\,II$\alpha$ photon energy along the $\overline{\Gamma}-\overline{N}$ direction of the Mn-derived surface Brillouin zone (SBZ) [see Fig.\,1(c,d)] for the Mn/W(110) system annealed at $T\approx300^\circ$\,C. The sharp PE features of the W(110) valence band~\cite{Dedkov:2008} (for example, at $\sim1.2$\,eV BE in the around normal-emission spectra) are easily detectable on top of broad Mn-derived features. These data confirm our previous conclusion that in this case high Mn-islands are formed. Fig.\,3(b) shows PE spectra measured in the same conditions for the Mn/W(110) system but annealed at a lower temperature of $T\approx200^\circ$\,C leading to the formation of a $(3\times3)$ overstructure in the LEED image. These spectra are characterized by two dispersion-less features: the first feature is in the vicinity of the Fermi level ($E_F$) and the second broad one at $\approx$ 2.5\,eV. According to the in-plane lattice constant of the bulk-like $\alpha$-Mn film extracted from the LEED images (9.21\,\AA) the border of the Mn-overlayer SBZ [$k_{||}=0.48$\,\AA$^{-1}$, $\overline{N}_{1/2,0}$-point, see Fig.\,1(c,d)] is reached at an emission angle of $\approx8^\circ$ at He\,II$\alpha$ photon energy. At this emission angle a slight decrease of the PE intensity in the vicinity of $E_F$ can be seen in the spectra and further it increases again when the emission angle approaches the center of the Mn-derived second SBZ, $\overline{\Gamma}_{1,0}$. In order to increase the resolution in $k$-space, an additional study of the electronic structure of the bulk-like $\alpha$-Mn(110) film was performed with He\,I$\alpha$ photon energy. The results are shown in Fig.\,3: (c) in the BE region between $E_F$ and 6\,eV and (d) in a zoomed region in the vicinity of the Fermi level. A clear variation of the PE intensity is observed at the Fermi level: it is relatively high at normal emission ($\overline{\Gamma}_{0,0}$-point in the first SBZ), decreases toward the border of the SBZ ($\overline{N}_{1/2,0}$-point), and increases again toward the center of the second SBZ ($\overline{\Gamma}_{1,0}$-point). An additional weak PE feature is resolved [see Fig.\,3(c)] in the second SBZ which disperses from $\approx3.3$\,eV up to $\approx4.8$\,eV BE.

In order to get more confidence in the interpretation of the observations, electronic structure calculations of non-magnetic \textit{bcc} $\alpha$-Mn and other competing phases were performed. Magnetism was disregarded, since the temperature during experiment was 300\,K, much larger than the magnetic ordering temperature $T_N=95$\,K. The program package CRYSTAL\,06~\cite{crystal06} was used. The gradient corrected functional of Perdew {\em et al.}~\cite{Perdew:1992} was employed. A $k$-point net with $12\times12\times12$ points was used. The chemically inactive [Ne] core of Mn was simulated by an energy-consistent full-relativistic pseudopotential: $\mathrm{Mn^{15+}}$-PP~\cite{Dolg:1987}. The basis set consists of  contracted Gaussian-type orbitals~\cite{basis}.

In Fig.\,4 the calculated total energies as functions of volume per Mn-atom are presented for the four Mn structural modifications ($\alpha$-, $\beta$-, $\gamma$-, and $\delta$-Mn). Experimental values for the internal parameters were used. In accordance with previous calculations~\cite{Hobbs:2003,Hafner:2003}, the  $\alpha$-phase is for all volumes the most stable Mn phase in the considered non-magnetic state. We attribute the deviations of the optimized lattice constants from the related experimentally values (for example, for $\alpha$-Mn: $a^{opt.}=8.56$\,\AA\ and $a^{expt.}=8.91$\,\AA~\cite{Yamada:1970} to the neglect of spontaneous magnetostriction.

In order to describe the experimental situation of a strained bulk-like $\alpha$-Mn film, we also performed calculations for orthorhombic Mn
(space group $Fmm2$). The in-plane lattice parameters were chosen to match the experimental strain, i.e., $c_{orth.}=a^{opt.}_\mathrm{\alpha-Mn}\times1.033=8.84$\,\AA\ and $b_{orth.}=a^{opt.}_\mathrm{\alpha-Mn}\times\sqrt{2}\times1.043=12.63$ \,\AA. The out-of-plane parameter $a_{orth.}$ was varied in order to mimick the relaxation toward zero out-of-plane stress and the internal parameters were fixed at their experimental values. The related energy vs. volume behavior is included in Fig.\,4. While this strained bulk-like film has as total energy somewhat higher than the ground state phase, its minimum energy is still below the energy of the other competing phases. It is thus very reasonable to assume that such a strained thick film can be (meta)stable under suitable conditions. 

Fig.\,5 shows calculated total density of states (DOS) and band structure along high symmetry lines in the Brillouin zone [see for notations Fig.\,1(c)] of the \textit{bcc} $\alpha$-Mn. The calculated DOS is in good agreement with previously published data~\cite{Hobbs:2003}.  The whole set of bands is rather dense as it is expected due to the large unit cell of $\alpha$-Mn, thus complicating any simple comparison between theory and experiment. Our experimental dispersions were measured in the $\overline{\Gamma}-\overline{N}$ direction of the $\alpha$-Mn(110) SBZ. Thus, the calculated electronic band dispersions along $\Gamma-N$, $\Gamma-H$, and $N-H$ directions of the bulk Brillouin zone of \textit{bcc} $\alpha$-Mn have to be considered in the analysis. The calculated band dispersions (Fig.\,5) reveal that in all three cases there are several bands in the vicinity of the Fermi level that disperse to $E_F$ when one goes from the center of the SBZ ($\overline{\Gamma}_{0,0}$-point) to the border ($\overline{N}_{1/2,0}$-point). This region is marked by shading. Assignment of the other PE features (non-dispersive band at 2.5\,eV BE and dispersive one in the region $3.3-4.8$\,eV BE) is complicated since many bands in the calculated band structure can contribute to the PE signal in this energy region. For example, in the band structure [Fig.\,5(b)] there is a series of almost non-dispersing electronic bands in the region of $1.5-3$\,eV BE that result in several maxima in the DOS. These bands can be assigned to the experimentally observed non-dispersing photoemission broad peak at 2.5\,eV BE. 

\textit{In conclusion,} we successfully stabilized bulk-like $\alpha$-Mn films with (110) orientation on W(110). The $(3\times3)$ overstructure observed for the Mn film with respect to the original W(110) LEED pattern is consistent with the suggested structural model. Density functional total energy calculations confirm that this structure, a strained $\alpha$-Mn phase, is close in energy to the ground state phase and thus can be prepared under carefully controlled conditions. It is metastable, though, and can be destroyed by annealing at $300 ^\circ$\,C. Angle-resolved PE spectra show weak but distinct dispersions of the electronic states in the valence band that confirm a high quality of the studied Mn films.

\newpage

\textbf{Figure captions:}
\newline
\newline
\textbf{Fig.\,1.}\,\,(Color online) (a) Crystalline structure of \textit{bcc} $\alpha$-Mn with 58 atoms in the cubic unit cell. (b) Possible arrangements of \textit{bcc} $\alpha$-Mn(110) and $\delta$-Mn(110) films on W(110). (c) Bulk and surface Brillouin zones with some high symmetry points of the \textit{bcc} lattice. (d) Reciprocal lattices of W(110) (dash-dotted line), $(3\times3)$\,Mn/W(110) (dashed line), and corresponding surface Brillouin zones of W(110) surface (thin solid line) and $(3\times3)$\,Mn/W(110) (thick solid line).
\newline
\newline
\textbf{Fig.\,2.}\,\,(Color online) Monitoring of the Mn growth on W(110). (a) XPS spectra of Mn/W(110): \#1- freshly deposited 50\,\AA\ thick Mn film on W(110), \#2- after annealing of \#1 at $T\approx200^\circ$\,C for 3 min, \#3- after annealing of \#1 at $T\approx300^\circ$\,C for 3 min, \#4- after annealing of freshly deposited 90\,\AA\ thick Mn film on W(110) at $T\approx200^\circ$\,C for 3 min, \#5- same as \#4 but annealed at $T\approx300^\circ$\,C for 3 min. LEED images of (b) W(110), (c) freshly deposited 50\,\AA\ thick Mn film on W(110) (corresponds to XPS spectrum \#1), (d) freshly deposited 50 or 90\,\AA\ thick Mn film on W(110) after annealing at $T\approx200^\circ$\,C for 3 min (corresponds to XPS spectra \#2 and \#4), and (e) same as (d) but annealed at $T\approx300^\circ$\,C for 3 min (corresponds to XPS spectra \#3 and \#5).
\newline
\newline
\textbf{Fig.\,3.}\,\,(Color online) Angle-resolved photoemission spectra of (a) freshly deposited 50 or 90\,\AA\ thick Mn film on W(110) after annealing at T$\approx300^\circ$\,C for 3 min [corresponds to XPS\#3 and \#5 in Fig.\,2(a) and LEED in Fig.\,2(e)] and (b,c) same as (a) but annealed at T$\approx200^\circ$\,C for 3 min [corresponds to XPS spectra \#2 and \#4 in Fig.\,2(a) and LEED in Fig.\,2(d)]. (d) Zoom of (c) for the energy region in the vicinity of the Fermi level. Spectra in (a,b) and (c,d) were measured with He\,II$\alpha$ (40.8\,eV) and He\,I$\alpha$ (21.2\,eV) photon energies, respectively.  
\newline
\newline
\textbf{Fig.\,4.}\,\,(Color online) Total energy per atom vs. volume per Mn-atom. Closed circles: $\alpha$-Mn; opened circles: $\beta$-Mn; triangles: $\gamma$-Mn; rombii: $\delta$-Mn and crosses: strained $\alpha$-Mn, see text.
\newline
\newline
\textbf{Fig.\,5.}\,\,(Color online) Total density of states as well as band structure of $\alpha$-Mn for the experimentally determined lattice constant ($a=9.21$\,\AA).


\begin{thebibliography}{10}
\bibitem{basis} The Mn basis set from Ref.~\cite{Dolg:1987} was employed with the innermost exponents being unchanged. The most diffuse exponents were reoptimized for the bulk yielding the values $0.17$, $0.17$, and $0.27$, for the $s$, $p$, and $d$-parts, respectively.
\end{thebibliography}
\end{document}